\documentclass[english,letterpaper,aps,prl,twocolumn,showpacs]{revtex4-1}
\usepackage{lmodern}

\usepackage[T1]{fontenc}
\usepackage[latin1]{inputenc}
\usepackage{graphicx}
\usepackage{amssymb}

\makeatletter

\usepackage{lmodern}
\usepackage{subfig}

\makeatletter

\usepackage{lmodern}

\makeatletter

\usepackage{lmodern}

\makeatletter

\usepackage{lmodern}

\makeatletter

\makeatletter

\usepackage{epsfig}

\usepackage{dcolumn}

\usepackage{bm}

\usepackage{bbm}

\usepackage{wasysym}

\newlength{\textwidthm}

\makeatother

\makeatother

\makeatother

\makeatother

\makeatother

\usepackage{babel}
\makeatother
\begin{document}

\title{Symmetry classification of energy bands in graphene and silicene}

\author{E. Kogan}
\email{Eugene.Kogan@biu.ac.il}
\affiliation{Department of Physics, Bar-Ilan University, Ramat-Gan 52900,
Israel}
\date{\today}

\begin{abstract}
We present the results of the symmetry classification of the electron energy bands in graphene and silicene using group theory algebra and the tight--binding approximation. The analysis is performed both in the absence and in the presence of the spin-orbit coupling. We also discuss the bands merging in the Brillouin zone symmetry points and the conditions for the latter to become Dirac points.
\end{abstract}


\maketitle

\section{Introduction}

Since graphene was first isolated experimentally \cite{novoselov}, it is in the focus of attention of both theorists and experimentalists. Obviously, understanding of the symmetries of the electrons dispersion law in graphene is of crucial importance. Actually, the symmetry classification of the energy bands in graphene (or "two-dimensional graphite") was presented nearly 60 years ago by Lomer in his seminal paper \cite{lomer}. Later the subject was analyzed by Slonczewski and Weiss \cite{slon}, Dresselhaus and Dresselhaus \cite{dresselhaus}, Bassani and Parravicini \cite{bassani}. Recent approaches to the problem are presented in the papers by Malard et al. \cite{malard}, Manes \cite{manes} and in our publication \cite{kogan}.

The present work has two aspects: a pragmatic and a pedagogical one. The first aspect is connected with the recent synthesis of  silicene, the counterpart of graphene for silicon, with buckled honeycomb geometry. This novel two-dimensional material has attracted considerable attention both theoretically \cite{hua,drummond} and experimentally recently, due to its exotic electronic structure and promising applications in nanoelectronics as well as its compatibility with current silicon-based electronic technology. So we present the symmetrty analysis of the silicene electron bands.

The pedagogical aspect is connected with the fact that different approaches to the symmetry classification, even if giving the same results, are based on different methods of applications of group theory. Thus in our previous paper [8] the labeling of the bands was based on compatibility relations and guesses. In the present work
we show that in the framework of the tight-binding approximation the representations of the little group in the symmetry points can be rigorously found in the framework of the group theory algebra. Though the idea of using the tight-binding approximation is by no means new (it was used already in the work by Lomer), our mathematical approach is totally different, as one can easily see comparing the present work with \cite{lomer}, and, to our opinion, more convenient for applications. This statement is supported by the analysis of the symmetry of the energy band in silicene.

We also generalize the symmetry classification by taking into account the spin-orbit coupling both for graphene and for silicene. This, to the best of our knowledge, wasn't done before even for graphene. Though in graphene the spin-orbit coupling is very weak, the problem is interesting in principle. One can expect that in silicenr the coupling is stronger, and it will become even more so for graphene related materials from heavier elements, provided they can be synthesized.

To remind to a reader a few basic things, important for the symmetry classification of the bands in any crystal, consider a point sub-group $R$  of the space group characterizing the symmetry of a crystal (we restrict ourselves with the consideration of symmorphic space groups). Any operation of the group $R$  (save the unit transformation) takes a general wavevector ${\bf k}$  into a distinct one. However, for some special choices of ${\bf k}$  some of the operations of the group $R$  will take ${\bf k}$  into itself rather than into a distinct wavevector. These particular operations are called the group of ${\bf k}$; it is a
subgroup of the group $R$. Points (lines) in the Brillouin zone for which the group of the wavevector contains elements other than the unit element are called symmetry points (lines). We may use a state (states) corresponding to such a special wavevector to generate a representation for the group of ${\bf k}$  \cite{kittel,harrison}. In this paper we consider crystals with the hexagonal Brillouin zone. In this case
the symmetry points are $\Gamma$  - the center of the Brillouin zone, the points $K$  which are corners of the Brillouin zone and the points $M$  which are the centers of the edges of the Brillouin zone.

\section{Tight--binding model}

We'll deal with the materials with a basis of two atoms per unit cell, and we'll search for the solution of Schroedinger equation as a linear combination of the functions
\begin{eqnarray}
\label{tb}
\psi_{\beta;{\bf k}}^j=\sum_{{\bf R}_j}\psi_{\beta}\left({\bf r}-{\bf R}_j\right) e^{i{\bf k\cdot R}_j},
\end{eqnarray}
where $\psi_\beta$ are atomic orbitals, $j=A,B$ labels the sub-lattices, and  ${\bf R}_j$ is the radius vector of an atom in the sublattice $j$.
A point symmetry transformation of the functions $ \psi_{\beta;{\bf k}}^j$ is a direct product of two transformations: the transformation of the sub-lattice functions $\phi^{A,B}_{{\bf k}}$, where
\begin{eqnarray}
\label{2}
\phi_{\bf k}^j=\sum_{{\bf R}_j} e^{i{\bf k\cdot R}_j},
\end{eqnarray}
and the transformation of the orbitals $\psi_{\beta}$. Thus the representations realized by the functions (\ref{tb}) will be the direct product of two representations.
Generally, this representation will be reducible. To decompose a reducible representation into the irreducible ones, it is convenient to use equation
\begin{eqnarray}
\label{ex}
a_{\alpha}=\frac{1}{g}\sum_G\chi(G)\chi_{\alpha}^*(G),
\end{eqnarray}
which shows how many times a given irreducible representation   is contained in a reducible one [11]. Additional information about the representations can be obtained if we use projection operator \cite{knox}
\begin{eqnarray}
O_{\alpha}=\frac{n_{\alpha}}{g}\sum_G\chi_{\alpha}^*(G)P(G),
\end{eqnarray}
where $n_{\alpha}$ is the dimensionality of the irreducible representation $\alpha$ and $P$ is the operator corresponding to a given transformation $G$ The operator projects a given function to the linear space of the representation $\alpha$. For a one dimensional representation the operator thus gives basis of the representation.

\section{Group theory analysis in the tight--binding model without the spin-orbit coupling}

Our tight-binding model space includes four atomic orbitals: $|s,p>$. (Notice that we assume only symmetry of the basis functions with respect to rotations and reflections; the question how these functions are connected with the atomic functions of the isolated carbon atom is irrelevant.)

\subsection{Graphene}

The Hamiltonian of graphene being symmetric with respect to reflection in the graphene plane, the bands built from the $|z>$  orbitals decouple from those built from the $|s,x,y>$  orbitals. The former are odd with respect to reflection, the latter are even. In other words, the former form $\pi$  bands, and the latter form $\sigma$  bands.

The group of wave vector ${\bf k}$  at the point $\Gamma$ is $D_{6h}$, at the point $K$  is $D_{3h}$ , at the lines $\Gamma-K$  is $C_{2v}$\cite{thomsen,kogan}. The group of wave vector ${\bf k}$  at the point $\Gamma$  is $D_{6h}$ , at the point $K$  is $D_{3h}$ , at the lines  $\Gamma-K$ is $C_{2v}$  \cite{thomsen,kogan}.
The  representations of the groups $D_{3h}$ and $D_{6h}$ can be obtained on the basis of identities
\begin{eqnarray}
D_{3h}=D_{3}\times C_s,\qquad D_{6h}= C_{6v}\times C_s.
\end{eqnarray}
The irreducible representations of the groups $D_{3}$ and $C_{6v}$ are presented in the Table 1. Each representation of the group, say $A_1$, begets two representations of the group $D_{3h}$ : $A_1'$  and $A_1''$; prime means that the representation is even with respect to reflection $\sigma_h$, double prime means that it is odd.

The irreducible representations of the group  $C_{6v}$  are  also presented in the Table 1. Because the inversion transformation $I$  can be presented as
\begin{eqnarray}
I=C_2\sigma_h,
\end{eqnarray}
the representations of the group   can be classified as symmetric (g) or antisymmetric (u) with respect to inversion.  Thus each representation of the group  $C_{6v}$, say $A_1$, begets two representations: $A_{1g}$  and $A_{1u}$.

\begin{table}
\begin{tabular}{|l|rrr|}
\hline
$D_3$ & $E$ & $2C_3$ & $3U_2$ \\
\hline
$A_1$ & 1 & 1 & 1   \\   $A_2$ & 1 & 1 & $-1$   \\  $E$ & 2 & $-1$ & 0   \\
\hline
\end{tabular}
\begin{tabular}{|l|rrrrrr|}
\hline
$C_{6v}$   & $E$ & $C_2$ & $2C_3$ & $2C_6$ & $3\sigma_v$ & $3\sigma_v'$ \\
\hline
$A_{1}$ &  1 & 1 & 1 & 1 & 1 & 1 \\
$A_{2}$ &  1 & 1 & 1 & 1 & $-1$ & $-1$ \\
$B_{2}$  & 1 & $-1$ & 1 & $-1$ & 1 & $-1$ \\
$B_{1}$  & 1 & $-1$ & 1 & $-1$ & $-1$ & 1 \\
$E_{2}$  & 2 & 2 & $-1$ & $-1$ & 0 & 0 \\
$E_{1}$  & 2 & $-2$  & $-1$ & 1 & 0 & 0 \\
\hline
\end{tabular}
\caption{Character table for irreducible representations of  $D_3$  and $C_{6v}$ point groups}
\label{table:d2}
\end{table}

Notice that the orbitals $|s>$  (or $|z>$ ) realize $A_1$  representation both of the group $D_3$  and of the group $C_{6v}$ , hence the representations of the groups
realized by the functions $\psi^{A,B}_{s,z;{\bf k}}$  will be identical to those  realized by the sub-lattice functions $\phi_{\bf k}^{A,B}$.

Let us start from the symmetry analysis at the point  $\Gamma$.
Because the transformations  $C_2,C_6,\sigma_v$   change sub-lattices, the characters corresponding to these transformations are equal to zero. The transformations  $E,C_3,\sigma_{v'}$  leave the sub-lattices as they were. Hence from Table \ref{table:d2} we see that the functions
$\phi_{\bf 0}^{A,B}$  realizes reducible representation
\begin{eqnarray}
R_{\Gamma}=A_1+B_2
\end{eqnarray}
of the group $C_{6v}$.

Taking into account the symmetry of the states relative to reflection in the plane of graphene , we obtain that at the point  $\Gamma$ the functions
$\psi_z$ (here and further on, when this is not  supposed to lead to a misunderstanding, we'll suppres the  index ${\bf k}$  in $\psi_{\beta;{\bf k}}$)  realize $A_{1u}$  and $B_{2g}$ representations of the group $D_{6h}$, characterizing $\pi$  bands, and the functions  $\psi_s$ realize  $A_{1g}$ and   $A_{2u}$ representations of the group, characterizing $\sigma$  bands.

Acting by projection operators $O_A$  and $O_B$  on a function $\psi^j$, we obtain that the irreducible representation $A_1$ is realized by symmetric combination of the  $A$ and $B$  orbitals, and the irreducible representation  $B$  - by the antisymmetric combination. One can expect that the first case occurs in the hole band, and the second - in the electron band.

The orbitals $|x,y>$  realize representation  $E_1$ of the group $C_{6v}$ \cite{landau}. Hence, representation of the group  realized by the  functions $\psi_{x,y}^{A,B}$  can be decomposed as
\begin{eqnarray}
E_1\times R_{\Gamma}=E_1+E_2.
\end{eqnarray}
Taking into account the symmetry of the states relative to reflection in the plane of graphene, we obtain that at the point $\Gamma$  the functions
$\psi_{x,y}$   realize $E_{1g}$  and $E_{2u}$  representations of the group $D_{6h}$, characterizing $\sigma$  bands.

To find wavefunctions realizing each of the irreducible representations we apply the projection operators and obtain
\begin{eqnarray}
O_{E_1}\psi^j_{\pm}\sim (\psi^j_{\pm}+\psi^{\overline{j}}_{\pm}),\quad O_{E_2}\psi^j_{\pm}\sim (\psi^j_{\pm}-\psi^{\overline{j}}_{\pm})
\end{eqnarray}
where  $\overline{j}=B$ if $j=a$, and vice versa.
Thus representation $E_1(E_2)$ is realized by symmetric (antisymmetric) combinations of $|x>$  and $|y>$  orbitals. One can expect that the first representation is realized at the hole band, and the second - at the valence band.

Now let us perform the symmetry analysis at the point $K$. The orbitals $|s>$  (or $|z>$ ) realize  $A_1$ representation of the group $D_3$.

The representation of the group realized by the functions $\phi^{A,B}_{\bf k}$  is determined by the transformation law of the exponentials $e^{i{\bf K\cdot R}_j}$  under the symmetry operations. Rotation of the radius vector by the angle $2\pi/3$  anticlockwise, is equivalent to rotation of the vector ${|bf K}$  in the opposite direction, that is to substitution of the three equivalent corners of the Brillouin zone:
${\bf K}\to {\bf K_2}\to {\bf K}_3\to {\bf K}$ where ${\bf K}=\left(2\pi/3a,2\pi/3\sqrt{3}a\right)$,
${\bf K}_2=\left(0,-4\pi/3\sqrt{3}a\right)$ and ${\bf K}_3=\left(-2\pi/3a,2\pi/3\sqrt{3}a\right)$.
The rotation multiplies each basis vector by the factor $e^{2\pi i/3}$. Using Eq. {\ref{ex}) we obtain
\begin{eqnarray}
a_{E}=\frac{1}{3}\left(2-e^{2\pi i/3}-e^{-2\pi i/3}\right)=1,
\end{eqnarray}
Hence the functions $\phi_{\bf K}^{A,B}$  realize irreducible representation $E$  of the group $D_3$.
Taking into account the symmetry of the states relative to reflection in the plane of graphene, we obtain representation   $E''$  of the group  $D_{3h}$, realized by the  $\psi_z$ functions, characterizing $\pi$  bands, and representation  $E'$ , realized by the $\psi_s$   functions, characterizing $\sigma$  bands.

Consider now representation of the group $D_3$  realized by the quartet of functions $\psi^j_{x,y}$  at the point  $K$. The orbitals $|x,y>$  realize representation $E$  of the group \cite{landau}. Hence, representation of the group  realized by the  functions $\psi_{x,y}^{A,B}$  can be decomposed as
\begin{eqnarray}
E\times E=A_1+A_2+E.
\end{eqnarray}
Taking into account the symmetry of the states relative to reflection in the plane of graphene, we obtain representations $A_1',A_2'$  and $E'$  of the group  $D_{3h}$, realized by the  $\psi_{x,y}$ functions, characterizing $\sigma$  bands.

Acting by projection operators, we obtain that the representation $A_1$  is realized by the vector space with the basis vector
$\psi^A_{+}+\psi^B_{-}$, and the representation  $A_2$  is realized by the vector space with the basis vector
$\psi^A_{-}-\psi^B_{+}$. The vector spaces realizing representations  $A_1$  and  $A_2$  being found, the representation   is obviously realized by the vector space spanned by the vectors  $\psi^A_+-\psi^B_-$, $\psi^A_-+\psi^B_+$.

Because the irreducible representation  $E'$ is realized both by $\psi_s$  and $\psi_{x,y}$  functions, these representations should be considered together. According to Wigner theorem \cite{petrashen} we still have two  representations, each of them being realized by two functions from a quartet $\psi_s^{A,B},\psi^A_+-\psi^B_-,\psi^A_-+\psi^B_+$.  Each $E'$ representation characterizes two $\sigma$  bands, merging at the point  $K$.
   The symmetry of the electron bands at the points $\Gamma$  and $K$  being determined, the symmetry at the lines $\Gamma-K$  follows unequivocally from the compatibility relations, presented in Table \cite{thomsen,kogan}.
\begin{table}
\begin{tabular}{|c|c|c|}
	\hline
$C_{2v}$ & $D_{6h}$ & $D_{3h}$  \\
\hline
Rep &\multicolumn{2}{c|} {Compatible with}  \\
	\hline
$A_{1}$   & $A_{1g},B_{2u},E_{1u},E_{2g}$ & $A_1',E'$  \\
$A_{2}$   & $A_{1u},B_{2g},E_{1g},E_{2u}$ & $A_1'',E''$   \\
$B_{1}$   & $B_{1u},A_{2g},E_{1u},E_{2g}$ & $A_2',E'$  \\
$B_{2}$   & $A_{2u},B_{1g},E_{1g},E_{2u}$ & $A_2'',E''$  \\
	\hline
\end{tabular}
\caption{Compatibility relations}
\label{table:c}
\end{table}
The table shows compatibility of the representations of the point group  , realized at the symmetry line  , with those realized at the symmetry points   and .
The results of this Section are presented on Fig. \ref{fig:bands}, reproduced from \cite{kogan}.
\begin{figure}[h]
\begin{center}
\centering
\includegraphics[width=0.45\textwidth]{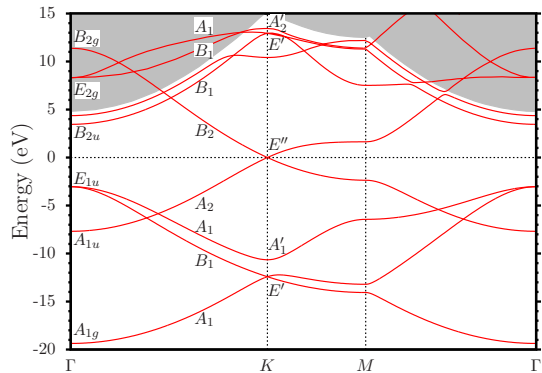}
\caption{\label{fig:bands}  (Color online) Graphene band structure evaluated with
use of the FP-LAPW method. The dashed line shows the Fermi energy. [Reproduced from Ref. \cite{kogan}.]}
\end{center}
\end{figure}

\subsection{Silicene}

The difference between silicene (or symmetrically equivalent to it buckled graphene) and graphene for our consideration is due solely to the decreased symmetry of the former. The group of the wavevector at the point $K$  in silicene is  $D_3$, at the point $\Gamma$  - $D_{3d}$  (this is also the point group of silicene). The representations of the
group $D_{3d}$  we can obtain on the basis of identity
\begin{eqnarray}
D_{3d}= D_3\times C_i.
\end{eqnarray}
 The direct product has twice as many representations as the group  $D_3$, half of them being symmetric (denoted by the suffix $g$), and the other half antisymmetric (suffix $u$) with respect to inversion. The characters of the representations of the group   are presented in the Table \ref{table:d3d}.
\begin{table}
\begin{tabular}{|l|rrrrrr|}
\hline
 Rep & $E$ & $2C_3$ & $3U_2$ & $I$ & $2S_6$ & $3\sigma_d$ \\
\hline
 $A_{1g}$ & 1 & 1 & 1 & 1 & 1 & 1 \\
  $A_{2g}$  & 1 & 1 & $-1$ & $1$ & $1$ & $-1$ \\
 $A_{1u}$ & 1 & 1 & $1$ & $-1$ & $-1$ & $-1$ \\
 $A_{2u}$ & 1 & 1 & $-1$ & $-1$ & $-1$ & 1 \\
 $E_{g}$ & 2 & $-1$ & 0 & 2 & $-1$ & 0 \\
 $E_u$ & 2 & $-1$  & 0 & $-2$ & 1 & 0 \\
\hline
\end{tabular}
\caption{Characters table for irreducible representations of   $D_{3d}$ point group}
\label{table:d3d}
\end{table}

The symmetry analysis in silicene  parallels that in graphene, so we'll be brief.

At the point $K$  the functions $\phi^{A,B}_{\bf K}$  realizes   irreducible representation $E$ of the group  $D_3$.

Orbitals $|s>$  and $|z>$  realize $A_1$  representation, and the orbitals $|x,y>$  realize $E$  representation of the group $D_3$. Thus at the point $K$  the functions $\psi_s$ realize representation $E$ of the group  $D_3$, and the same same can be said about the functions  $\psi_z$. Reducible representation realized by the functions $\psi_{x,y}$ can be decomposed into the irreducible ones:
\begin{eqnarray}
E\times E=A_1+A_2+E.
\end{eqnarray}
So when the symmetry is reduced by going from graphene to silicene, the representations $A_1'$  and $A_2'$  turn into $A_1$  and $A_2$. Representation  $E''$ and two representations $E'$  turn into three representations $E$. Loosing the reflection in plane symmetry, we can not claim now that one representation  is realized exclusively by  $|z>$ orbitals. All the $E$  representations mix $|s,p>$  orbitals.

At the point  $\Gamma$ the functions  $\phi^{A,B}_{{\bf 0}}$   realizes reducible representation of the group  $D_3$:
\begin{eqnarray}
R_{\Gamma}=A_{1g}+A_{2u}.
\end{eqnarray}
Orbitals $|s>$  realize $A_{1u}$, orbitals  $|z>$ -- $A_{2u}$, and  orbitals $|x,y>$ -- $E_u$
representations of the group.
Thus the wavefunctions (\ref{tb}) realize reducible representation of the group $D_{3d}$ which is
\begin{eqnarray}
&&(A_{1u}+A_{2u}+E_u)\times(A_{1g}+A_{2u})\nonumber\\
&&=A_{1u}+A_{2g}+A_{2u}+A_{1g}+E_g+E_u.
\end{eqnarray}
So when the symmetry is reduced by going from graphene to silicene, the representations   $E_{2g}$ and $E_{1u}$ turn into the representations $E_g$ and $E_u$ respectively.

The band structure of silicene is being different from that of graphene, the
merging of the bands  is no different. The statement becomes clear when comparing Fig. \ref{fig:drummond}, reproduced from Ref. \cite{drummond},  with Fig. \ref{fig:bands}.
\begin{figure}[h]
\begin{center}
\centering
\includegraphics[width=0.45\textwidth]{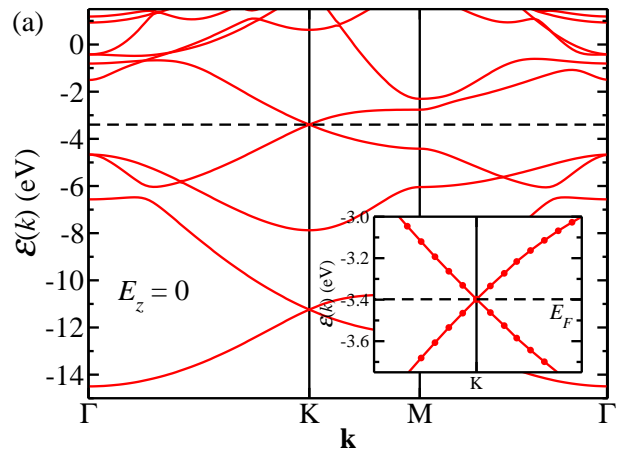}
\caption{\label{fig:drummond}  (Color online) DFT-PBE band structures for silicene. The dashed line shows the Fermi energy  and the
insets show the spectrum near the Fermi level in the vicinity of the K
point. [Reproduced from Ref. \cite{drummond} with permission.] }
\end{center}
\end{figure}

\section{Group theory analysis in the tight-binding model with the spin--orbit coupling }

In the absence of spin-orbit coupling, electron spin can be taken into account in a trivial way: each band we considered was doubly spin degenerate.
When the spin-orbit coupling is taken into account, the algebra, actually, is not that different. The symmetry, and the representations realized by the sub-lattice functions (\ref{2}) remaim the same. However, instead of atomic orbitals we should consider atomic terms: $\beta$  enumerates states terms from doublets $|s>^{(1/2)}$, $|p>^{(1/2)}$  and quartet $|p>^{(3/2)}$.

Due to the semi-integer value of $J$  we have to consider double-valued representations realized by the atomic terms (and by the crystal wave functions). To find (the characters of) two-valued irreducible representations, it is convenient to introduce the concept of a new element of the group (denoted by  $Q$); this is a rotation through an angle  $2\pi$ about an arbitrary axis, and is not the unit element, but gives the latter when applied twice:  $Q^2=E$.

The characters of the rotation by angle $\phi$  applied to the term $|\dots>^{(j)}$  is
\begin{eqnarray}
 \chi_j(\phi)=\frac{\sin\left(J+\frac{1}{2}\right)\phi}{\sin\frac{1}{2}\phi}.
\end{eqnarray}
With respect to the inversion $I$ the character is
 \begin{eqnarray}
 \chi_j(I)=\pm(2J+1),
 \end{eqnarray}
where the sign plus corresponds to the $s$ states, and the sign minus to the $p$ states. Finally, the characters corresponding to reflection in a plane $\sigma$  and rotary
reflection through an angle $\phi$  are found writing these symmetry transformations as
\begin{eqnarray}
\sigma=I C_2,\qquad S(\phi)=IC(\phi+\pi).
 \end{eqnarray}
Both in graphene and in silicene we'll restrict ourselves by  the symmetry analysis at the point $K$.

\subsection{Graphene}

The sub-lattice functions  $\phi^{A,B}_{\bf k}$ realize $E_g$ representation of point group $D_{3h}$. The electron terms realize two-valued representations of the group, which are presented in Table \ref{table:ll} \cite{koster}.
\begin{table}
\begin{tabular}{|l|ccccccccc|}
\hline
 $D_{3h}'$   & $E$ & $Q$ & $\sigma_h$ & $C_3$ & $C_3^2$ & $S_3$ & $S_3^2$ & $3U_2$ & $3\sigma_v$ \\
  &  &   & $\sigma_hQ$ & $C_3^2Q$ & $C_3Q$ & $S_3^2Q$ & $S_3Q$ & $3U_2Q$ & $3\sigma_vQ$ \\
  \hline
 $\Gamma_7$  & 2 & $-2$ & 0 & 1 & $-1$ & $-\sqrt{3}$ & $\sqrt{3}$ & 0 & 0 \\
 $\Gamma_8$  & 2 & $-2$ & 0 & 1 & $-1$ & $\sqrt{3}$ & $-\sqrt{3}$ & 0 & 0 \\
 $\Gamma_9$  & 2 & $-2$ & 0 & $-2$ & 2 & 0 & 0 & 0 & 0 \\
\hline
\end{tabular}
\caption{Characters for two-valued irreducible representations of group $D_{3h}'$}
\label{table:ll}
\end{table}

Doublet $|s>^{(1/2)}$ realizes $\Gamma_7$  representation of the group; doublet $|p>^{(1/2)}$ realizes  $\Gamma_8$ representation, quartet  $\psi_{p^{(3/2})}$  realizes  $\Gamma_7$ representation twice.  (We decided to use chemical notation for the single-valued representation, and BSW notation for double-valued representations [\cite{kittel}].)  The sub-lattice functions realize representation  $E_g$  of the group; from Eq. (\ref{ex}) we obtain
\begin{eqnarray}
\Gamma_7\times E_g=\Gamma_8\times E_g=\Gamma_7+\Gamma_8.
\end{eqnarray}
Thus at the point $K$  four bands realize representation  $\Gamma_7$ of the group  each, and four bands realize representation $\Gamma_8$  each. In particular, we obtained the (well known) result that the four-fold degeneracy (including spin) of the bands merging at the point $K$  is partially removed by the spin-orbit coupling, and only two-fold (Kramers) degeneracy is left.

\subsection{Silicene}

The two-valued representations of   are presented in Table \ref{table:ll2} \cite{koster}.
\begin{table}
\begin{tabular}{|l|rrrrrr|}
\hline
 Rep   & $E$ & $Q$  & $C_3$ & $C_3^2$ &  $3U_2$ & $3U_2Q$ \\
  & &  & $C_3^2Q$ & $C_3Q$ & & \\
  \hline
$\Gamma_5$  & 1 & $-1$ & $-1$ & 1 & $i$ & $-i$  \\
$\Gamma_6$  & 1 & $-1$ & $-1$ & 1 & $i$ & $-i$ \\
 $\Gamma_4$  & 2 & $-2$ & $1$ & $-1$ & $0$ & $0$ \\
\hline
\end{tabular}
\caption{Characters for two-valued irreducible representations of group $D_{3}'$}
\label{table:ll2}
\end{table}

Each of the doublets $|s>^{(1/2)},|p>^{(1/2)}$  realizes $\Gamma_4$  representation of the group. Quartet  $|p>^{(3/2)}$  realizes this representation twice. The sub-lattice functions realize representation $E$  of the group. From Equation (\ref{ex}) we obtain
\begin{eqnarray}
\Gamma_4\times E=\Gamma_4+\Gamma_5+\Gamma_6.
\end{eqnarray}
(For the same reasons as for ordinary representations, two complex conjugate two-valued representations   must be regarded as one physically irreducible representation of twice the dimension.)
Thus at the point $K$  four bands  realize representation $\Gamma_4$  of the group $D_3'$ each, and four bands realize representation $\Gamma_5+\Gamma_6$  each.

\section{Dirac points}

In this final part of the paper we would like to clarify the relation between the symmetry and the existence of Dirac points.
According to the classical approach \cite{hund,herring}, the merging of the bands at a point ${\bf k}_0$  is connected with the multi (higher than one) - dimensional representation of the space group $G_0$, realized in this point. Looking for a linear dispersion point in the vicinity of the merging point we may use the degenerate ${\bf k\cdot p}$  perturbation theory. Let a two-dimensional irreducible representation is realized at a point ${\bf k}_0$. Expanding the wavefunction with respect to the basis of the two-dimensional irreducible representation
 \begin{eqnarray}
\psi({\bf k})=\sum_{i=1}^2c_i({\bf k})\psi_i({\bf k}_0),
\end{eqnarray}
for the expansion coefficients we obtain equation
\begin{eqnarray}
\label{dirac}
\sum_{j=1}^2\frac{{\bf k\cdot p}_{ij}}{m}c_j({\bf k})=\epsilon({\bf k}) c_i({\bf k}),
\end{eqnarray}
where ${\bf p}_{ij}=<\psi_i({\bf k}_0)|\hat{\bf p}|\psi_j({\bf k}_0)>$ (of course, we need the absence of inversion symmetry at the point, for the matrix elements to be different from zero).
The dispersion law is given by the equation
\begin{eqnarray}
\label{dispersion}
\epsilon({\bf k})=\sum_{\alpha}a_{\alpha}k_{\alpha}\pm\sqrt{\sum_{\alpha\beta}\gamma_{\alpha\beta}k_{\alpha}k_{\beta}},
\end{eqnarray}
 where $\alpha,\beta$ are cartesian indexes $x,y$. Eq. (\ref{dispersion}) should contain only combinations of wavevector components which are invariant
 with respect to all elements of the group $G_0$.
 In the case when the group $G_0$ does not have any vector invariants, and the only tensor invariant is the quantity $k_x^2+k_y^2$,
  we obtain the dispersion law
 \begin{eqnarray}
 \epsilon({\bf k})=\pm vk,
 \end{eqnarray}
which, like it was shown by Dirac himself in 1928, guaranties that Eq. (\ref{dirac}) is Dirac equation, in the sense the the matrices $p_x$ and $p_y$ satisfy anticommutation relations
\begin{eqnarray}
p_x^2=p_y^2\propto I,\qquad \{p_x,p_y\}_+=0,
\end{eqnarray}
where $I$ is the unity matrix.

To be more specific, consider the groups of wavevector at the point $\Gamma$;  in graphene  it is $D_{3h}$,
and  in silicene it is $D_{3}$.
In both cases, to find the dispersion law at the point $\Gamma$ it is enough to study invariants of the group $D_3$.
And we can easily check up that both conditions, necessary for the existence of the Dirac point, are satisfied.
And we can easily check up that both conditions, necessary for the existence of the Dirac point, are satisfied. This explains, in particular, why the band calculations show the existence of Dirac points in silicene \cite{hua,drummond}, which has  lower symmetry than graphene.

In general, the role of the tight binding approximation in symmetry classification of the bands in graphene, like its role in symmetry classification of bands
in other crystals,  is only auxiliary. The approximation greatly helps in the classification and sheds additional light on the nature of the bands. but one must remember that
that there are more important things that this or that approximation - and that is symmetry.

\section{Conclusions}

This paper presents an applications of group theory to  very important cases of graphene and silicene.

\section{Acknowledgments}

Discussions with J. L. Manes, V. Falko and Hua Jiang were very illuminating for the author.


\begin{thebibliography}{99}


\bibitem{novoselov} K. S. Novoselov, A. K. Geim, S. V. Morozov, D. Jiang,
Y. Zhang, S. V. Dubonos, I. V. Grigorieva, A. A. Firsov, Science {\bf 306}, 5696 (2004).

\bibitem{lomer} W. M. Lomer, Proc. Roy. Soc. A {\bf 227}, 330 (1955).

\bibitem{slon}J. C. Slonczewski and P. R. Weiss,  Phys. Rev. {\bf 109}, 272 (1958).


\bibitem{dresselhaus} G. Dresselhaus and M. S. Dresselhaus, Phys. Rev. {\bf 140}, A401 (1965).

\bibitem{bassani} F. Bassani and G. Pastori Parravicini, Nuovo Cim. B {\bf 50}, 95 (1967).

\bibitem{malard} L. M. Malard, M. H. D. Guimaraes, D. L. Mafra, M. S. C. Mazzoni, and A. Jorio, Phys. Rev. B {\bf 79}, 125426 (2009).

\bibitem{manes} J. L. Manes, Phys. Rev. B  {\bf 85}, 155118 (2012).

\bibitem{kogan} E. Kogan and V. U. Nazarov, Phys. Rev. B {\bf 85}, 115418 (2012).

\bibitem{hua} C.-C. Liu, Hua Jiang, Yugui Yao, \prb {\bf 84}, 195430 (2011).

\bibitem{drummond} N. D. Drummond, V. Zolyomi, and V. I. Falko, \prb {\bf 85}, 075423 (2012).

\bibitem{kittel} C. Kittel, {\it Quantum Theory of Solids}, (John Wiley \& and Sons. Inc. New--York, London 1963).

\bibitem{harrison} W. A. Harrison, {\it Solid State Theory}, (McGraw Hill Book Company, New York--London--Toronto, 1970).

\bibitem{landau} L. D. Landau and E. M. Lifshitz, {\it Quantum Mechanics}, (Pergamon Press, 1991).

\bibitem{knox} R. S. Knox and S. Gold, {\it Symmmetry in the Solid State}, (W. A. Benjamin, Inc., 1964, New York, Amsterdam).

\bibitem{thomsen}  C. Thomsen,  S. Reich,  J. Maultzsch, {\it Carbon Nanotubes: Basic Concepts and Physical Properties}, (Wiley Online Library,
2004 WILEY-VCH Verlag GmbH).

\bibitem{petrashen} M. I. Petrashen, E. D. Trifonov, J. L. Trifonov, {\it Applications of Group Theory in Quantum Mechanics} (Dover Publications,  2009).

\bibitem{koster} G.F. Koster, J.O. Dimmock, R.G. Wheeler, and H. Statz, {\it Properties of the Thirty-Two Point Groups} (MIT Press, Cambridge 1964).

\bibitem{hund} F. Hund, Z. Physik {\bf 99}, 119 (1936).

\bibitem{herring} C. Herring, \prb {\bf 52}, 365 (1937); J. Franklin Inst. {\bf 223}, 525 (1942).

\end{thebibliography}
\end{document}